# Improving Estimation of Portfolio Risk Using New Statistical Factors[*]

Xialu Liu[1], John Guerard[2], Rong Chen[3], and Ruey Tsay[4]

[1]Department of Management Information systems, San Diego State University, San Diego, CA 92182

[2]Independent Researcher, Bluffton, SC 29910

[3]Department of Statistics, Rutgers University, Piscataway, NJ 08854

[4]Booth School of Business, University of Chicago, Chicago, IL 60637

## Abstract

Searching for new effective risk factors on stock returns is an important research topic in asset pricing. Factor modeling is an active research topic in statistics and econometrics, with many new advances. However, these new methods have not been fully utilized in asset pricing application. In this paper, we adopt the factor models, especially matrix factor models in various forms, to construct new statistical factors that explain the variation of stock returns. Furthermore, we evaluate the contribution of these statistical factors beyond the existing factors available in the asset pricing literature. To demonstrate the power of the new factors, U.S. monthly stock data are analyzed, and the partial $F$ test and double selection LASSO method are conducted. The results show that the new statistical factors bring additional information and add explanatory power in asset pricing. Our method opens a new direction for portfolio managers to seek additional risk factors to improve the estimation of portfolio returns.

[*] Ruey Tsay is the corresponding author. Address: 5807 South Woodlawn Avenue, Chicago, IL 60637. Email: ruey.tsay@chicagobooth.edu. The authors wish to thank Guanhao Feng and Xin He for sharing their code to download and analyze data. Chen's research was supported in part by National Science Foundation grants DMS-2027855, DMS-2052949 and DMS-2319260.
ETHICAL STATEMENTS: The authors have no competing interests to declare that are relevant to the content of this article. This article does not contain any studies with human participants performed by any of the authors.

**Keywords**: Asset pricing, Fama-French portfolios, Matrix factor models, Principal component analysis, Portfolio selection, Statistical factors.

## 1. Introduction

The Efficient Frontier, as conceived by Markowitz (1952), created much of modern portfolio selection; see for example Markowitz (1959); Guerard et al. (2013, 2014, 2015); Geczy et al. (2020); Geczy and Guerard (2023). The Efficient Frontier sought to produce the highest expected portfolio return, for a given level of risk, or the least amount of risk for a given level of return. The reader is referred to Markowitz (1952, 1959), Bloch et al. (1993), and Markowitz et al. (2021). The Efficient Frontier concept, its estimation and implementation, was rigorously discussed by Harry Markowitz for both academicians and practitioners in Bloch et al. (1993). That is, one must estimate models of expected returns and covariance matrices, and often employ portfolio optimization techniques, initially produced in Markowitz (1956). The reader is referred to Guerard et al. (2021) and Guerard et al. (2024) for an updated collection of applied investment research analyses, including several by Harry Markowitz and his various co-authors. Harry Markowitz shared the Nobel Prize in Economic Sciences in 1990 with William F. (Bill) Sharpe. Bill Sharpe enhanced the Efficient Frontier analysis by initially estimating a single index model, whose stock return regression coefficient on the market return variable would become known as "beta". The independently developed capital market equilibrium theories of Sharpe (1964), Li and Li (1996), and Mossin (1966) became known as the Capital Asset Pricing Model, CAPM. Beta represented a single risk factor, a measure of systematic risk. Sharpe processed to enhance his CAPM estimations by developing multi-factor models of risk, see Sharpe (2012). In the early 1990s, Eugene Fama and Ken French collaborated on a series of papers (Fama and French, 1992, 2008, 2016) that sought to identify factors associated with portfolio selection. The reader is referred to Cochrane (2017) for the selected papers of Eugene Fama. An adequate literature review of portfolio selection would encompass far more pages than this issue to honor Harry Markowitz. However, the goal of our analysis is to seek a more properly estimated portfolio risk using improved statistical analysis of covariances that build upon the Fama-French methodology.

The multi-factor models have attracted lots of attention in the recent literature. Readers are referred to Feng et al. (2020) for the explosion in the number of risk factors and for ways to test the potential contribution of new factors. These factors are constructed from domain knowledge by finance experts for modeling portfolio risks. On the other hand, there also exist many statistical factor models in which the factors are estimated from the data under study. From King (1966), Stone (1970, 1974), to Stone and Guerard (2010), researchers have used the CRSP data to search for portfolio risk factors to enhance portfolio Sharpe Ratios. For example, the well-known principal component analysis (PCA) and its variants are often used to produce statistical factors. See, for instance, Jolliffe (2002). These statistical factors are derived by statistical methods mainly for dimension reduction rather than for modeling risks. They are often used to summarize the information embedded in the observed data and the analysis is classified as unsupervised learning in the modern machine learning literature. It is then useful to synthesize financial factors and statistical factors. In particular, under the current big data environment, a portfolio may consist of a large number of assets so that the estimation of covariance for portfolio selection becomes problematic, especially when the sample size is not sufficiently larger than the number of assets.

In this article, we construct several new statistical factors based on matrix factor models, provide a comprehensive analysis to evaluate the contribution of these statistical factors, and to explore the relationships between financial factors and these new statistical factors. A key concept used in our analysis is to exploit the rich data structure to construct informative statistical factors. Specifically, we construct a new set of statistical factors derived from two matrix factor models and demonstrate the power of these factors in further explaining the variation of stock returns on top of the traditional Fama-French factors as well as the vast existing financial factors. With a comprehensive data analysis, we are able to show that the newly constructed statistical factors can make significant contribution beyond the Fama and French three-factor models and the momentum factor in explaining financial risks. The significance is shown by statistical tests.

The rest of the paper is organized as follows. Section 2 reviews portfolio risks and the existing factors used in risk analysis in portfolio management, and introduces new factors based on matrix factor models in Tucker form and CP (CANDECOMP/PARAFAC) form. Section 3 presents the results of empirical analysis to demonstrate the power of the new factors by evaluating their contribution in explaining the stock returns over the traditional Fama-French factors. Section 4 shows that the new factors also make significant contribution against many other existing factors

(the factor zoo) using the double selection (DS) LASSO method of Feng et al. (2020). Section 5 concludes.

## 2. Portfolio Risk and New Statistical Factors

Understanding the sources of portfolio risks plays a critical role in portfolio management. The knowledge can be used to mitigate potential risk by hedging or asset reallocation. Under the market efficient hypothesis (Fama, 1970), stock prices (or returns) are believed to contain all information of the assets and can be used to extract valuable information pertaining to portfolio risks. For example, the market factor is thought to represent market volatility and market condition associated with the status of the economy, and the momentum factor is used to describe the local dynamic trend. Those well-known financial factors are built by domain knowledge. However, the sources of financial risk cannot be fully explained by the existing financial factors. In other words, the existing factors, no matter how many of them, cannot explain fully the variations embedded in stock returns. We believe that properly extracted statistical factors that make use of the return structure can be useful in modeling portfolio risks. To this end, we start with a brief review of the existing factors before introducing the proposed new statistical factors.

### 2.1 A brief review of existing factors

In this section, we briefly review some available factors constructed in the finance literature. To explain the average excess returns on the U.S. stocks and empirical abnormalities over the fundamental CAPM model, Fama and French (1993) identified three stock-market common factors, the overall market factor, the firm size premium factor, and the book-to-market equity premium factor. The model, commonly known as the Fama-French three-factor model, can be written as

$$R_i(t) - RF(t) = a_i + b_i[RM(t) - RF(t)] + s_i SMB(t) + h_i HML(t) + e_i(t), \quad (1)$$

where $R_i(t)$ is the return of the $i$-th portfolio at time $t$, $RF(t)$ is the risk free rate of return at time $t$, $RM(t)$ is the market return at time $t$ (often using the S&P 500 index return as proxy), $SMB(t)$ is the size premium factor at time $t$, $HML(t)$ is the value premium factor at time $t$, and $e_i(t)$ is the error term, for $t = 1, \ldots, T$, and $i = 1, \ldots, n$. In addition, $a_i$, $b_i$, $s_i$, and $h_i$ are factor coefficients for the $i$-th portfolio, $n$ is the number of portfolios, and $T$ is the length of time period (or the sample size). Fama and French (1993) constructs the size factor SMB in (1) by sorting all stocks by their capital values and obtaining the difference between returns of the portfolios formed by the largest 50\% stocks and the smallest 50% stocks. Similarly, the value factor is constructed by sorting all stocks by their book-to-market equity (BE/ME) values and obtaining the difference between returns of the portfolios formed by the top 30% stocks and the bottom 30% stocks.

To evaluate the effectiveness of the constructed factors, Fama and French (1993) estimates model (2) using 100 ($10 \times 10$) test portfolios formed on size and book-to-market equity and shows that the factors are statistically significant in explaining additional variation beyond the standard CAPM model. Since then, many new factors were constructed and found to be useful, including the momentum factor (Carhart, 1997; Fama and French, 2012) and many others (Abarbanell and Bushee, 1998; Hirshleifer et al., 2004; Rechardson et al., 2010).

The Fama-French factors are constructed based on the domain knowledge of finance experts. They are weighted averages of returns of all stocks in the market, with fixed and specifically designed weights depending on certain characteristics (e.g. rank by the size or BE/ME ratio) of individual stocks. Although these designed factors have been shown to be effective in various important applications, including portfolio management, risk management, and others, there is room for further improvement by constructing additional effective factors using statistical approaches to determining the optimal weighting schemes.

Statistical factors and their associated factor models are completely data driven and do not require any domain knowledge. The dynamic multivariate factor model usually puts the observed data at time $t$ in a vector $\boldsymbol{x}_t$ to facilitate the description of contemporary relationship between different time series. It assumes that

$$\boldsymbol{x}_t = \boldsymbol{A}\boldsymbol{f}_t + \boldsymbol{\epsilon}_t, \quad t = 1, \ldots, T, \tag{2}$$

where $\{\boldsymbol{x}_t\}$ is an observed $n \times 1$ process, $\boldsymbol{A}$ is an $n \times r$ loading matrix, $\{\boldsymbol{f}_t\}$ is an $r$-dimensional latent factor process, $\boldsymbol{\epsilon}_t$ is the error term at time $t$ and $r \ll n$. A special feature of Model (2) is that both the loading matrix $\boldsymbol{A}$ and the factor process $\boldsymbol{f}_t$ are unobserved and need to be estimated. In addition, $r$ also requires estimation and $\boldsymbol{f}_t$ can and usually does accommodate certain serial dependence to explain the temporal relationship of the observed process.

One of the goals of factor analysis is dimension reduction. When the dimension of time series $n$ is very large, traditional time series models, including the commonly used vector autoregressive moving-average model, may suffer from over-parametrization (Lütkepohl, 2015) and need more computational resources and time for proper analysis. Model (2) reduces the dimension in the sense that the $n$-dimensional process is driven by the $r$-dimensional common factors. This provides a possible solution for solving the over-parametrization problem.

There are various methods developed for estimating the latent factor model in (2) (Forni et al., 2000; Sentana and Fiorentini, 2001; Bai and Ng, 2002; Stone and Guerard, 2010; Lam et al., 2011; Chang et al., 2015; Su and Wang, 2017; Gao and Tsay, 2022, 2024+) and its inference procedures (Bai, 2003; Onatski, 2010; Lam and Yao, 2012; Chen et al., 2014; Barigozzi et al., 2018; Chen et al., 2020) in the literature. The multivariate factor analysis has been widely applied in different fields, including finance (Chang et al., 2015; Liu and Chen, 2016; Massacci, 2017; Barigozzi et al., 2018; Liu and Chen, 2020; Liu and Zhang, 2022; Pelger and Xiong, 2022), economics (Stock and Watson, 2002a,b, 2009; Chen et al., 2014; Ma and Su, 2018) and environmental sciences (Pan and Yao, 2008; Lam and Yao, 2012; Gao and Tsay, 2019).

Most of the estimation methods employ the PCA technique and its variations. For example, in the econometrics literature, Stock and Watson (2002b) applied PCA to obtain the diffusion indices for forecasting. Statistical properties of diffusion indices have been studied by Bai and Ng (2002). These authors also propose methods to determine the number of important factors.

PCA was developed for dimension reduction of multivariate data in the form of a vector. On the other hand, with the advances in data collection and prior knowledge of the data under study, modern big data often contain certain structures. In particular, many financial data are available naturally in a matrix form. For example, all public companies publish their financial statistics annually, hence forming an annually observed matrix time series. Option data is also naturally in a matrix form with maturity and strike prices as column and rows, respectively. Although by stacking the elements in a matrix into a long vector, traditional PCA analysis can be used to reduce dimension and form factors. However, such an approach disregards the row and column classifications which often contain significant information. The stacking approach also significantly increases the dimension hence often requires significantly larger number of parameters in the estimation, leading to less accurate results.

Wang et al. (2019) and Chen et al. (2020) proposed a factor model in Tucker form for matrix time series and Han et al. (2024) proposed a factor model in the CP form, also for matrix time series. These models fully utilize the matrix structure of the data in a bilinear form, hence achieving significant dimension reduction and producing more stable and accurate results.

In this paper, we will use Fama-French $10 \times 10$ portfolio returns formed on Size and Book-to-Market as the basis to construct new statistical factors. Although one can vectorize these 100 portfolio returns to form a 100-dimensional vector of returns and perform PCA to extract statistical factors, such an analysis, however, overlooks the data structures concerning size and book-to-market. To maintain the data structure of a matrix, we employ matrix factor models. Two new sets of statistical factors are to be constructed, using the matrix factor model in Tucker form and in CP form.

There are several reasons of using of Fama-French 10 by 10 portfolio as the base of our factor construction, instead of directly using individual stocks with their numerical size, BE/ME ratio and other financial information. First, it has been recognized that size and BE/ME are two most important abnormality of the CAPM model and contains significant information that can be used to guide portfolio management and risk management. Second, individual stock returns are often too noisy with large variation, which leads to unstable and inaccurate factor construction. By sorting stocks by their size and BE/ME value and constructing portfolios with each decile of size and BE/ME, portfolios are formed with significant variation reduction, which will lead to more stable and accurate factor construction. Third, by forming the $10 \times 10$ portfolio returns in the matrix form, size and value classification structure are retained. Instead of forming size factor and value factor separately as Fama and French proposed, the matrix approach allows us to take into account of the correlation between the size and value at various level, hence providing effective factor construction.

## 2.2 Factors constructed by matrix factor models in Tucker form

The matrix factor model in Tucker form was first introduced by Wang et al. (2019), see also Chen et al. (2020); Han et al. (2023); Chen and Fan (2023); Liu and Chen (2022) and Gao and Tsay (2023). Specifically, it assumes that the data observed are structured in a matrix form and there are two loading matrices to compress the rows and columns of the observed process separately. Let $X_t$ denote a matrix time series observed at t of size $n_1 \times n_2$. The model assumes that

$$X_t = RF_tC' + E_t, t = 1, \dots, T, \qquad (3)$$

where $F_t$ is an $r_1 \times r_2$ ($r_1 \ll n_1$ and $r_2 \ll n_2$) latent matrix process with time varying latent common factors, $R$ is an $n_1 \times r_1$ front loading matrix, $C$ is an $n_2 \times r_2$ back loading matrix, and $E_t$ is an $n_1 \times n_2$ error matrix. The dimension reduction is achieved in the sense that the $n_1 \times n_2$ process is driven by $F_t$ with a much smaller size.

It is well known that the matrix factor model in Tucker form suffers from a rotational ambiguity issue. Specifically, for Model (3), the terms $R$, $C$, and $F_t$ can be replaced with $RU_1$, $CU_2$ and $U_1^{-1}F_t U_2^{-1}$ for any non-singular $r_1 \times r_1$ matrix $U_1$ and $r_2 \times r_2$ matrix $U_2$. However, the linear spaces spanned by the columns of $R$ and $C$, called the row loading space and column loading spaces are uniquely defined. Hence one often aims to estimate a representative of the equivalent class of estimates. Since rotation ambiguity is linear in nature, any representative can be used in any downstream linear analysis without problem.

Chen et al. (2022) assumes that the error process in (3) is white but with arbitrary contemporary correlation, and the factor process contains all dynamics in the $X_t$ process. Accordingly, they propose two estimation procedures, TOPUP and TIPUP, using lagged co-moments and eigen-analysis. The estimated representative of the loading matrices $R$ and $C$ are orthonormal. Han et al. (2023) introduces an iterative procedure to improve these estimators with faster convergence rates. Chen and Fan (2023) and Gao and Tsay (2022) assume that the factors explain most of the contemporary correlation in $X_t$ hence requires the error $E_t$ has weak contemporary correlation and can be weakly serially correlated. Once the orthonormal estimates of $\widehat{R}$ and $\widehat{C}$ are obtained, the factors can be obtained with

$$\widehat{F}_t = \widehat{R}' X_t \widehat{C}.$$

We treat $\{\widehat{F}_t\}$ as factors constructed by matrix factor model in Tucker form.

## 2.3 Factors constructed by matrix factor models in CP form

Han et al. (2024) introduces a matrix factor model in the CP form. Specifically, let $X_t$ be a matrix of dimension $n_1 \times n_2$. The matrix factor model in CP form assumes

$$X_t = \sum_{i=1}^{r} f_{it} a_i b_i' + E_t, \ t = 1, \dots, T, \tag{4}$$

where $a_i$ and $b_i$ are $n_1$-dimension and $n_2$-dimension vectors, respectively, with $||a_i|| = ||b_i|| = 1$ for $i = 1, \dots, r$. The time varying latent factors $f_{it}, i = 1, \dots, r$ are assumed to be uncorrelated processes, and $E_t$ is assumed to be a white noise process, but with contemporary correlations among the elements in $E_t$. Note that the model assumes that the signal part of the process $X_t$ is a weighted sum of rank-one matrices $a_i b_i'$, with time varying weights serving as latent factors. The rank-one matrices provide clear indication of the relationships between columns and rows.

Han et al. (2024) proposed an estimation procedure based on a composite PCA initiation based on lagged co-moments of $X_t$ (to utilize the white noise structure) and an iterative simultaneous orthogonalization procedure that utilizes the uncorrelated factors assumption. Theoretical guarantee was investigated of the performance of estimation procedure.

Once we obtain the estimated loading vectors $\widehat{a}_i$ and $\widehat{b}_i$, the factor process can be estimated by

$$\hat{f}_{it} = \tilde{\boldsymbol{a}}_i' X_t \, \tilde{\boldsymbol{b}}_i, \quad i = 1, \dots, r,$$

where $\tilde{\boldsymbol{a}}_i'$ is $\hat{\boldsymbol{a}}_i$ projected onto the space orthogonal to the space spanned by $\{\hat{\boldsymbol{a}}_j, j = 1, i-1, i+1, \dots, r\}$ and is obtained from the final iteration of the iterative estimation algorithm. $\tilde{\boldsymbol{b}}_i$ is defined similarly. The estimated factors $\{\hat{f}_{it}, i = 1, \dots, r\}$ are called CP factors in this paper.

**Remark 1.** The estimation procedures for constructing the Tucker and CP factors are computational efficient since they are based on the PCA approach, instead of computationally intensive optimization procedures. For all financial modelling exercises, an important practical issue is to determine how much historical data should be used in the modeling, as market structure does change over time. Using only recent historical data limits the sample size, hence the estimation accuracy, while using long historical data runs into the problem of potential structure change and model instability. On the other hand, factors constructed statistically have the advantage for adapting structure changes, hence providing more useful factors for portfolio managers. Through backfitting exercises, it is possible to be adaptive in determining the amount of historical data to be used by treating it as a tuning parameter. For the matrix factor models used, an important tuning parameter is the number of factors to be used. Using too many factors may result in model overfitting while too few factors may miss the important market information. General statistical model selection procedures such as those proposed in Han et al. (2022) may be used. For financial applications, backfitting is an effective tool to achieve a proper balance.

## 3. Empirical Analysis

In this section, we assess the contribution statistical factors make beyond the factors of Fama and French (1993) and the momentum factor. In the following we refer the market, size, BE/ME, and momentum factors collectively as Fama-French factors. We use the regression models with Fama-French factors as controls and include the new statistical factors of interest. $F$ tests are conducted to evaluate if the coefficients of statistical factors are significantly different from zero.

## 3.1 Data and baseline model for comparison

We downloaded monthly individual stock data from January 2000 to December 2017 from Wharton Research Data Services. To remove the impact of the financial crisis, observations between January 2007 and December 2009 were excluded. Hence, the sample size is $T = 16 \times 12 = 192$. The data set includes all NYSE, Amex and NASDAQ stocks traded during the sampling period. The number of stocks is $n = 10278$. Fama-French factors were obtained from French's website, https: //mba.tuck.dartmouth.edu/pages/faculty/ken.french/data_library.html. The new statistical matrix factors were estimated with the returns of $10 \times 10$ portfolios formed on size and BE/ME. One thing worth noting is there are many missing values of the returns of 100 portfolios. Thus, for the regression models, the sample size may be different. To make sure each regression model has sufficient data for making reliable inference, we only analyze the stocks that have more than 30 observations. There are 4720 stocks used for modeling.

We use matrix factor models in Tucker form introduced in Wang et al. (2019) and in CP form in Han et al. (2024) to estimate the statistical factors. As discussed in Remark 1, the number of factors to be used is an important tuning parameter in matrix factor models. In this exercise, we use $2 \times 2$ for the Tucker factors, using the rank range provided by the statistical approaches, then adjust it according to empirical performance in a back-fitting exercise. For fair comparison, the number of CP factors is also set to be 4. Estimation is done using iterative TIPUP procedure of Chen et al. (2022) for Tucker factors and the estimation procedure in Han et al. (2024) for the CP factors.

We consider the following regression models for an individual stock across *t*,

$$R_i(t) - RF(t) = a_i + b_i[RM(t) - RF(t)] + s_i SMB(t) + h_i HML(t) + m_i MOM(t) + e_i(t), \quad (5)$$

$$R_i(t) - RF(t) = a_i + b_i[RM(t) - RF(t)] + s_i SMB(t) + h_i HML(t) + m_i MOM(t) + \boldsymbol{\beta}_i' \boldsymbol{f}(t) + e_i(t), \quad (6)$$

where $\boldsymbol{\beta}_i$ in (6) is a *K*-dimensional vector, and $\boldsymbol{f}(t)$ is the *K*-dimensional constructed statistical factors. For Tucker factors, $\boldsymbol{F}_t$ is a $2 \times 2$ matrix hence $K = 2 \times 2 = 4$. For CP-factors, $K = r =$

4. Model (5) is the **reduced model** including Fama-French factors and the momentum factor only, and Model (6) is the **full model** with all factors.

## 3.2 The power of the constructed Tucker factors

### 3.2.1 Explanatory analysis

Table 1 report the estimated loading matrices of a matrix factor model in Tucker form. The estimated $R$ indicates that all size portfolios load on the first row of $F_t$ with similar weights (the first column of $\widehat{R}$), while the loadings on the second row of $F_t$ show a U-shaped pattern, from positive to negative, as the size increases. The estimated $C$ has similar patterns.[1]

Table 1. Estimated loading matrices for Tucker factors

|    | $\widehat{R}$ | | $\widehat{C}$ | |
|----|-------|--------|-------|--------|
| 1  | 0.444 | 0.515  | 0.449 | 0.450  |
| 2  | 0.382 | 0.274  | 0.416 | 0.380  |
| 3  | 0.368 | 0.237  | 0.374 | 0.070  |
| 4  | 0.326 | 0.025  | 0.296 | 0.140  |
| 5  | 0.277 | -0.085 | 0.321 | -0.112 |
| 6  | 0.235 | -0.139 | 0.276 | -0.211 |
| 7  | 0.249 | -0.247 | 0.260 | -0.278 |
| 8  | 0.271 | -0.224 | 0.270 | -0.337 |
| 9  | 0.270 | -0.317 | 0.208 | -0.363 |
| 10 | 0.273 | -0.603 | 0.189 | -0.499 |

---

[1] Note that if both $R$ and $C$ are in the form that the first column is constant 1, and in the second column the first five elements are 1 and the next five elements are -1 (a contrast vector), then with proper normalizing, the four constructed factors are (i) the average of all 100 portfolio returns (similar to the market factor), (ii) the **difference** of the average returns of all small size portfolios and that of all big size portfolios (similar to Fama-French's SMB factor), (iii) the difference of the average returns of all low BE/ME portfolios and that of all high BE/ME portfolios (similar to Fama-French's HML factor), and (iv) the difference of (a) the average return difference of small size and low BE/ME portfolios and big size and low BE/ME portfolios, and (b) the average return difference of small size and high BE/ME portfolios and big size and high BE/ME portfolios. The last factor is the difference of differences, reflecting the interaction between size and BE/ME, a unique feature of jointly analyzing size and BE/ME. The constructed Tucker factors use optimally determined weights (especially the U-shaped weights), instead of using *ad hoc* weights.

Table 2 Descriptive statistics and correlation between Fama-French factors and Tucker factors

| Variable | Ex ret[a] | Market | SMB | HML | MOM | TF1[b] | TF2 | TF3 | TF4 |
|---|---|---|---|---|---|---|---|---|---|
| Ex ret | - | - | - | - | - | - | - | - | - |
| Market | 0.47 | - | - | - | - | - | - | - | - |
| SMB | 0.37 | 0.44 | - | - | - | - | - | - | - |
| HML | 0.04 | 0.12 | 0.15 | - | - | - | - | - | - |
| MOM | -0.18 | -0.09 | 0.03 | -0.30 | - | - | - | - | - |
| TF1 | 0.49 | **0.94** | **0.71** | 0.21 | -0.09 | - | - | - | - |
| TF2 | 0.32 | 0.63 | 0.46 | **0.80** | -0.25 | **0.70** | - | - | - |
| TF3 | 0.19 | **0.72** | -0.18 | 0.09 | -0.09 | 0.49 | 0.43 | - | - |
| TF4 | 0.27 | 0.59 | 0.28 | 0.54 | -0.27 | 0.62 | 0.67 | 0.40 | - |
| Mean | -0.06 | 0.94 | 0.01 | 0.03 | 0.49 | 9.86 | 1.71 | 3.09 | -0.01 |
| Standard deviation | 0.16 | 3.39 | 2.24 | 2.09 | 2.87 | 42.11 | 11.78 | 9.03 | 5.06 |
| VIF | - | 179.76 | 57.78 | 16.25 | 1.14 | 315.07 | 24.11 | 7.80 | 4.27 |

a. Ex ret denotes the excess return of an individual stock, the left-hand side of (5) and 6).
b. TF stands for Tucker factor.

Note: The correlation is written in bold if its absolution value is higher than or equal to 0.7.

Table 2 reports the descriptive statistics and correlations of variables. It can be seen that the first Tucker factor is highly correlated with the market factor with correlation coefficient 0.94. This is because the loadings in the first columns of $\widehat{R}$ and $\widehat{C}$ are similar, hence all 100 portfolios were used to construct the first factor with similar weights. It is not surprising since by construction the first factor has the highest explaining power of the return variability while the market factor is known to be the dominant factor in stock markets. On the other hand, the first factor is also markedly related to the SMB factor, so it is not a replacement of the market factor, due to the estimated varying weights. We can regard the first Tucker factor as a weighted index of the market and size effect. The second Tucker factor is mainly related to the HML factor so that it provides a proxy for the value premium. Overall, Table 2 shows that the first three Tucker factors capture most of the variability explained by the three Fama-French factors. We calculated the VIF values of the independent variables including the Fama-French factors and the Tucker factors in the full model and present the results in the last row of Table 2. It confirms that there exist multi-collinearities when all factors are included in Model (6). To address this problem and obtain numerically stable results, we regressed the Tucker factors on Fama-French factors and the momentum factor and replace the Tucker factors with the resulting residuals in Model (6). These residuals mark the new contribution of the Tucker factors.

### 3.2.2 Results of Tucker factors

We fit models (5) and (6) using all 4720 individual stock returns as the response variable. Here the estimated Tucker factors are used as ft in (6). For each stock $i$, we obtain the $R^2$ of the estimated models. Table 3 summarizes these $R^2$ for the two models, and Figure 1 plots the histograms of these $R^2$ values. Form the table and figure, we see that the selected Tucker factors increase substantially the $R^2$ of the fit.

Table 3. Summary of $R^2$ comparison with and without the selected Tucker factors in the model over all stocks. The full model includes the Tucker factors.

|        | Reduced model (5) | Full model (6) |
|--------|-------------------|----------------|
| Mean   | 0.271             | 0.401          |
| Median | 0.247             | 0.410          |

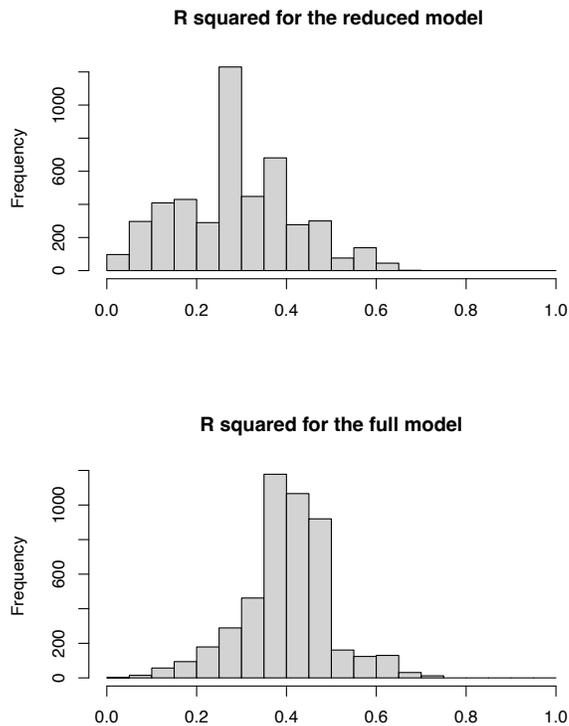

Figure 1: Histograms of $R^2$ of the regression models with and without the Tucker factors in the model. The full model includes the selected Tucker factors.

In order to confirm that the increases in $R^2$ are statistically significant, next we conduct the partial $F$ test to check whether the regression coefficients of the Tucker factors are not zero. Figure 2 shows the histogram of $p$-values of these tests. Among them, 68.9% of $p$-values are less than 0.05 and 72.4% of $p$-values are less than 0.1. The partial $F$ test confirms that the Tucker factors contribute significantly to explaining the return variability beyond that of the Fama-French factors.

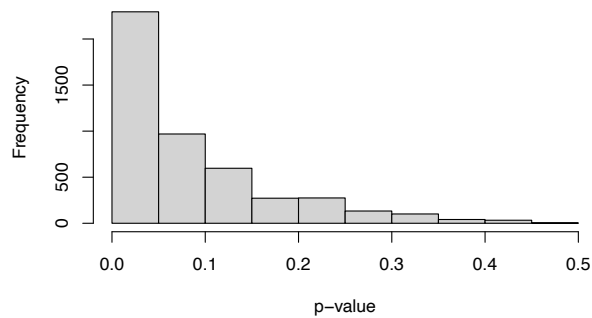

Figure 2: Histogram of $p$-values for the partial $F$ test when the selected Tucker factors are included in the full model.

## 3.3 The power of the constructed CP factors

Now we turn to the constructed CP factors. We basically perform a similar analysis by replacing the Tucker factors by the CP factors.

### 3.3.1 Explanatory analysis

Table 4 reports the estimated $\widetilde{a}_i$ and $\widetilde{b}_i$ of the matrix factor model in CP form. It helps us understand how CP factors were constructed. For each $\widetilde{a}_i$ and $\widetilde{b}_i$ if we only consider the three largest elements in absolute value, the portfolios with small/medium size and large BE/ME values load heavily on the first factor, portfolios with very small/very large size and small/medium BE/ME values load heavily on the second factor, portfolios with large size and all BE/ME values load heavily on the third factor (similar to the SMB factor, but only with large size without the difference), and portfolios with all sizes and medium/large BE/ME values load heavily on the forth factor (similar to the HML factor, but only with high BE/ME without the difference). They are quite different from the Tucker factors and are less interpretable.

Table 4. Estimation results for CP factors

|    | $\tilde{a}_1$ | $\tilde{b}_1$ | $\tilde{a}_2$ | $\tilde{b}_2$ | $\tilde{a}_3$ | $\tilde{b}_3$ | $\tilde{a}_4$ | $\tilde{b}_4$ |
|----|-------|-------|--------|--------|--------|-------|--------|--------|
| 1  | 0.198 | 0.084 | 0.656  | -0.120 | -0.157 | 0.358 | 0.414  | 0.159  |
| 2  | -0.207| 0.199 | 0.327  | -0.195 | 0.102  | 0.013 | -0.058 | 0.310  |
| 3  | 0.424 | 0.054 | 0.064  | 0.586  | -0.024 | 0.530 | -0.017 | -0.027 |
| 4  | 0.082 | 0.145 | -0.287 | 0.100  | -0.065 | 0.069 | 0.651  | -0.013 |
| 5  | 0.447 | 0.178 | -0.161 | 0.244  | 0.076  | 0.317 | 0.183  | 0.398  |
| 6  | 0.444 | 0.188 | 0.079  | 0.398  | 0.092  | 0.351 | 0.373  | 0.441  |
| 7  | 0.335 | 0.684 | 0.016  | 0.457  | 0.338  | 0.081 | 0.105  | -0.010 |
| 8  | 0.347 | 0.168 | 0.243  | 0.385  | 0.288  | 0.257 | 0.254  | 0.701  |
| 9  | 0.234 | 0.525 | -0.037 | -0.068 | 0.333  | 0.430 | 0.365  | 0.030  |
| 10 | 0.219 | 0.302 | -0.532 | 0.123  | 0.799  | 0.323 | 0.140  | 0.181  |

Table 5. Descriptive statistics and correlation between Fama-French factors and Tucker factors

| Variable | Ex ret[a] | Market | SMB | HML | MOM | CP1[b] | CP2 | CP3 | CP4 |
|----------|--------|--------|------|------|------|------|------|------|------|
| Ex ret   | -      | -      | -    | -    | -    | -    | -    | -    | -    |
| Market   | 0.47   | -      | -    | -    | -    | -    | -    | -    | -    |
| SMB      | 0.37   | 0.44   | -    | -    | -    | -    | -    | -    | -    |
| HML      | 0.04   | 0.12   | 0.15 | -    | -    | -    | -    | -    | -    |
| MOM      | -0.18  | -0.09  | 0.03 | -0.30| -    | -    | -    | -    | -    |
| CP1      | -0.39  | -0.66  | -0.24| -0.16| -0.06| -    | -    | -    | -    |
| CP2      | 0.05   | 0.15   | 0.14 | -0.07| -0.12| -0.12| -    | -    | -    |
| CP3      | 0.43   | **0.84** | 0.58 | 0.41 | -0.12| 0.64 | 0.08 | -    | -    |
| CP4      | 0.45   | **0.79** | 0.59 | 0.23 | -0.05| -0.46| 0.10 | **0.79** | - |
| Mean     | -0.06  | 0.94   | 0.01 | 0.03 | 0.49 | -1.99| 0.86 | 1.79 | 1.67 |
| Standard deviation | 0.16 | 3.39 | 2.24 | 2.09 | 2.87 | 4.76 | 4.13 | 11.84 | 6.78 |
| VIF      | -      | 6.18   | 1.87 | 1.75 | 1.21 | 2.09 | 1.82 | 7.25 | 3.62 |

a. Ex ret denotes the excess return of an individual stock, the left-hand side of (5) and 6).
b. CP stands for CP factor.

Note: The correlation is written in bold if its absolution value is higher than or equal to 0.7.

The sample correlations among the estimate CP factors and Fama-French factors are shown in Table 5. From the table, we see that the 3-rd and 4-th CP factors are highly correlated with the market factor and the SMB factor. In particular, the third CP factor also has marked correlation with the HML factor. It seems that the third and fourth CP factors are linear functions of the Fama-French factors. The first CP factor is also a function of the Fama-French factors. The second CP factor is not highly correlated with the Fama-French factors, indicating that this particular CP factor represents a new dimension not described by the Fama-French factors. The VIF values for all factors (as explanatory variables in the full model) are moderate.

Table 6. Summary of $R^2$ comparison with and without the selected CP factors in the model over all stocks. The full model includes the CP factors.

|   | Reduced model (5) | Full model (6) |
|---|---|---|
| Mean | 0.292 | 0.356 |
| Median | 0.290 | 0.354 |

## 4. The power of the constructed factors vs the factor zoo

Hundreds of factors were proposed to explain the cross section of stock returns (Abarbanell and Bushee, 1998; Hirshleifer et al., 2004; Rechardson et al., 2010; Fama and French, 2016; Feng et al., 2018) as noted by Harvey et al. (2016) and by Feng et al. (2020). In this section, we consider adding more factors in the regression model (6) as controls to evaluate the additional contribution of the Tucker factors and CP factors beyond the existing factors (the factor zoo as called by Feng et al. (2020). When the number of controls is too large, it may result in poor estimation and invalid inference due to the curse of dimensionality. Here we adopt the DS LASSO method employed by Feng et al. (2020) to select the useful factors.

Specifically, let $r_t$ be an observed $n \times 1$ vector of portfolio returns. Let $h_t$ be a $p \times 1$ high-dimensional existing potentially useful vectors, and $g_t$ be an $r \times 1$ vector of new factors to be tested. Feng et al. (2020) considers the following model for the expected returns

$$E(r_t) = \iota_n \gamma_0 + C_g \lambda_g + C_h \lambda_h, \tag{7}$$

where $\iota_n$ is an $n \times 1$ vector of 1s, $C_a = Cov(r_t, a_t)$ for $a = g$ or $h$. The significance of the new factors is determined by the test whether the corresponding stochastic discount factor loading, $\lambda_g$, is significantly different from zero.

In this study, the factor candidate set $h_t$ contains 135 factors with $p = 135$ proposed in literature before 2012. Feng et al. (2020) constructed 750 test portfolios from these 135 factors with $n = 750$. The complete list of these factors can be found in Appendix of Feng et al. (2020) and more details of these portfolios can be seen in Section II of Feng et al. (2020). In our analysis, we consider the four new Tucker factors or the four CP factors, with $r = 4$. We used the data set provided by Feng et al. (2020) which contains monthly returns of 750 portfolios from July 1976 to December 2017.

To evaluate if any new factors bring additional valuable information, we would like to test if the $4 \times 1$ vector $\lambda_g = 0$ is significantly different from zero. DS LASSO yields an estimator for $\lambda_g$ and the standard error of this estimator. We conducted the Wald test to check if all the components of $\lambda_g$ are zero. The $p$-values, reported in Table 7, show that the Tucker factors make significant contribution after being added to the model on explanation of the stock returns and the contribution of CP factors are only marginal significant at the conventional 5% level. Overall, we found the newly constructed Tucker factors and CP factors make significant contribution beyond the existing factor zoo.

Table 7. $p$-values of Wald Test in Model (7)

|  | $p$-value |
|---|---|
| Tucker factors | 0.041 |
| CP factors | 0.063 |

**Remark 2**. The newly constructed statistical factors may potentially replace the Fama-French factors in portfolio management tools, or as additional factors in the pool for risk evaluations. Since these factors are tradable as they are linear combinations of the Fama-French 10-by-10 portfolios (with the estimated weights shown in Table 1 and Table 4), they can be used in hedging strategies, in constructing α-portfolios and in other quantitative trading strategies.

**Remark 3**. Harry Markowitz was interested in the interactions of risk factor models and the variables in the expected return models. In this study, we have demonstrated empirically that the statistical factors enhance risk models. It is beyond the scope of this current analysis as to the expected portfolio enhancements of the better risk models, including transaction costs. It would be interesting to investigate the effectiveness of the 135 factors in Feng et al. (2020) and the proposed new statistical factors versus a much smaller stock selection models, say 10 factors, as in Markowitz et al. (2021). We leave this important issue for future research.

## 5. Summary

This paper introduced new statistical factors derived from advanced matrix factor models, specifically in the Tucker and CP forms, to improve portfolio risk estimation. By integrating these factors into existing asset pricing models, the study demonstrates a significant improvement in explanatory power over traditional factors, such as those found in the Fama-French model. The empirical analysis reveals that these new factors are statistically significant, offering superior risk estimation, particularly in scenarios involving complex and large datasets. The results indicate that incorporating these statistical factors into portfolio management strategies can substantially enhance decision-making and overall portfolio performance. This research presents a novel approach to risk management with important implications for both academic research and practical application in finance.

Our empirical study showed that including the new statistical factors in the Fama-French model improves the model fitting performance for most of the individual stocks, hence these factors will be useful factors in improving the evaluation of portfolio risks. In addition, we also showed empirically the new statistical factors are useful against the factor zoo in providing portfolio risk estimations. Harry Markowitz, in Bloch et al. (1993) and Markowitz et al. (2021), always held that models of expected returns and covariances were necessary inputs to portfolio selection. The Tukey-Bisquare robust regression used in Bloch et al. (1993) continues to produce statistically significant models for expected returns for portfolio selection, 30 years after publication, see Guerard et al. (2024).

Statistical factors, by definition, are constructed based on data alone (in our case, return data). Their economic meanings are difficult to comprehend. One may attempt to match the constructed factors with observed economic series, such as the expected inflation, monthly growth in industrial production, interest rate differentials, and changes in the term structure of interest rates (Chen et al., 1986), hence provide certain economical findings. But it is in general difficult as these statistical factors tend to be combinations of multiple economic forces. Note that the proposed factors are based on Fama-French 10x10 portfolios, hence they are only revealing the risk factors related to size and BE/ME ratio. They are more powerful than the Fama-French factors (at least empirically) because it explores the interaction between size and BE/ME ratio, and the combination weights are optimally determined, instead of in the ad hoc way the Fama-French factors are constructed. The method is different from the existing literature in constructing statistical factors as it is the first method that utilizes the matrix time series structure of Fama-French $10 \times 10$ portfolios, with shown empirical power.

Multi-factor risk models utilizing covariance estimation techniques can effectively help asset managers control risk for portfolio selection. Although these new factors cannot replace the traditional financial factors such as the Fama-French factors, they provide a new direction for portfolio managers to seek additional information in assessing financial risks. As the statistical factors adjust the construction weights dynamically based on the current return data, they are more adaptive to the changes of market conditions.